\documentclass[final,3p,times,twocolumn]{elsarticle}
 \biboptions{comma,sort&compress}

\usepackage{graphicx}
\usepackage{here}

\def\beq{\begin{equation}}
\def\eeq{\end{equation}}
\def\bea{\begin{eqnarray}}
\def\eea{\end{eqnarray}}
\def\bq{\begin{quote}}
\def\eq{\end{quote}}

\def\nnb{\nonumber}
\def\ga{\left(}
\def\dr{\right)}

\def\lrar              {\Longrightarrow}
																
\def\nnb{\nonumber}
\def\la{\langle}
\def\ra{\rangle}
\def\nin{\noindent}
\def\ba{\vspace*{-0.2cm}\begin{array}}
\def\ea{\end{array}\vspace*{-0.2cm}}

\def\b{$\bullet~$}
\def\als{\alpha_s}

\def\gg2{ \la\alpha_s G^2 \ra}
\def\gg3{g^3f_{abc}\la G^aG^bG^c \ra}
\def\ggg4{\la\als^2G^4\ra}

\journal{Nuc. Phys. (Proc. Suppl.)}

\begin{document}

\begin{frontmatter}

\title{SVZ sum rules : 30 $\oplus$ 1 years later} 

 \author[label1]{Stephan Narison\corref{cor1} }
   \address[label1]{Laboratoire
Univers et Particules, CNRS-IN2P3,  
Case 070, Place Eug\`ene
Bataillon, 34095 - Montpellier Cedex 05, France.}
\ead{snarison@yahoo.fr}


\pagestyle{myheadings}
\markright{ }
\begin{abstract}
\noindent
For this exceptional  25th anniversary of the QCD-Montpellier series of
conferences initiated in 85 with the name ``Non-perturbative methods", we take the opportunuity
to celebrate the 30 $\oplus$ 1 years of the discovery of the SVZ (also called ITEP, QCD or QCD spectral)
sum rules by M.A. Shifman, A.I. Vainshtein and V.I. Zakahrov in 79~\cite{SVZ}. In this talk, I have the duty to present the status of the method. I shall (can) not enumerate the vast area of successful applications of  sum rules in hadron physics but I shall focus on the historical evolution of field and its new developments. More detailed related discussions and more complete references can be found in the textbooks \cite{SNB1,SNB2}. 
 \end{abstract}

\end{frontmatter}
\section{Introduction}
\vspace*{-0.25cm}
 \nin
This talk, which is supposed to be a status review, aims to complete the contributions of the two inventors (M.A. Shifman and V.I. Zakahrov), the historical talk of H.G. Dosch on baryon sum rules and the talks given by diffferent orators on various modern applications of the sum rules in this session. This ceremonial session is chaired by E. de Rafael who has participated with enthousiasm in the developments of the field. \\
To my opinion, the SVZ sum rule \cite{SVZ} is one of the most important discovery of the 20th century in high energy physics phenomenology \cite{ZAK,SHIF,DOSCH, CONF,ZAKA,BERTL,RRY,NOV,SHUR,IOFFE,REV,SNB1,SNB2}.
This discovery has been recognized by the award of the Sakurai price to SVZ in 1999. \\
This year is the 25th annivesary of the QCD Montpellier series of conferences where S and Z participate regularly and where Z belongs as a committee member during several years. Then, it looks a natural recognition of their efforts and works to also celebrate, in the same time, the 30 $\oplus$ 1 years of their discovery, also because the chairman of this conference works intensively in this field since its invention. \\
The SVZ ideas were fantastic as they have formulated in QCD, with the inclusion of the non-perturbative condensate contributions, the old idea of (ad hoc) duality \cite{DUALITY}, superconvergent \cite{WEINBERG} and smearing \cite{POGGIO} phenomenological sum rules   used in the era of 60-76.   In the same time, some attempts to improve these pre-QCD sum rules using perturbative QCD  have been however done \cite{FNR} using finite energy sum rules (FESR) contour techniques \`a la Shankar \cite{SHANKAR} for testing the breaking of the convergence of the Weinberg sum rules  \cite{WEINBERG} when the perturbative current quark masses are switch on. As a consequence, some combinations of superconvergent sum rules for non-zero light quark masses have been  proposed. 
\vspace*{-0.5cm}
\section{Different forms of the sum rules}
\vspace*{-0.5cm}
 \nin
From these short historical reminders, the SVZ sum rules are, like previous others, alternative  improvements of the well-known K\"allen-Lehmann dispersion relation, which for a hadronic two-point correlator  reads:
\bea
\Pi_{H}(Q^2)&\equiv &i\int d^4x~e^{iqx}\la 0| {\cal T} J_H(x) J_H^\dagger (0)|0\ra
\nnb\\
&=&\int_{t<}^{\infty}{dt\over t+Q^2+i\epsilon}{1\over\pi}{\rm Im} \Pi(t)+...
\eea
where $Q^2\equiv -q^2>0$, ... means arbitrary subtraction terms which are polynomial in t; $J_H(x)$ is any hadronic current with definite quantum numbers built from 
quarks and/or gluon fields. This well-known dispersion relation is very important as it relates $\Pi(Q^2)$ which can be calculated in QCD provided that $Q^2$ is much larger than the QCD scale $\Lambda^2$, with its imaginary part which can be measured at low energy from experiments. 
\vspace*{-0.5cm}
\subsection{The SVZ sum rules}
\vspace*{-0.25cm}
 \nin
SVZ improvements  act in the two sides of this dispersion relation. 
\subsubsection*{\b Exponential  sum rules}
\nin
The {\it popular sum rule} is obtained  when one  takes an infinite number of derivatives $n$ of the correlator in $Q^2$ but keeping the ratio $Q^2/n\equiv M^2\equiv \tau^{-1}$ (Borel/Laplace sum rule scale) fixed. In this way, one can eliminate the substraction terms and the dispersion becomes an exponential:
\beq
 {\cal L}(\tau)= \int_{t_<}^{\infty} dt~{e}^{-t\tau}~{1\over\pi}{\rm Im} \Pi(t)~,
 \nnb
\vspace*{-0.25cm}
\eeq
where the exponential has the nice feature to enhance the lowest resonance contribution in the spectral integral. Another related sum rule is the ratio of sum rules:
\beq
 {\cal R}(\tau)\equiv -{d\over d\tau}\log {\cal L}(\tau)~,
 \label{eq:ratio}
 \eeq
which is often used in the literature for extracting the lowest ground state hadron mass as  at the optimization point (disussed later on) $ {\cal R}(\tau_0)\simeq M_R^2$.
\vspace*{-0.25cm}
\subsubsection*{\b Moment sum rule }
\nin
An alternative sum rule, which possesses the same property of enhancing the lowest resonance contribution is the moment sum rule: 
\bea
 {\cal M}_n&\equiv&{(-1)^n\over n!}{ d^n\over (dQ^2)^n}\Pi(Q^2)
 \nnb\\
 &=&\int_{t_<}^{\infty} {dt\over(t+Q^2)^n} ~{1\over\pi}{\rm Im} \Pi(t) ,
\vspace*{-0.25cm}
\eea
which has been (first discussed) to my knowledge by Yndurain using the positivity of the spectral function \cite{27} and has been applied later on to light quarks \cite{43,LEL,SNB1} and to heavy quark systems \cite{SVZ,NOV,RRY}. 
\subsection{Some other QCD spectral sum rules (QSSR)}
 \nin
Different variants of the SVZ 
sum rules have been proposed in the literature such as the  \footnote{We shall not discuss the analytic continuation
and infinite norm approaches \cite{ANALYTIC,MENES}, based on a $\chi^2$-minimization fitting procedure in the complex $q^2$-plane, where some comments on this approach have been given in \cite{SNB1}.}:
\vspace*{-0.25cm}
\subsubsection*{\b Finite energy sum  rule (FESR)}
\nin
This sum rule  \cite{DUALITY,FNR,PEROTTET}:
\beq
{\cal M}_n\equiv\int_{t_<}^{t_c} {dt~ t^n} ~{1\over\pi}{\rm Im} \Pi(t) ~~~~n> 0~,
\eeq
can derived from the Laplace/Borel sum rule in the limit $\tau\to 0$. This FESR is often called local duality (in the contrast to the above called global duality)
sum rule demonstrates the correlation between the lowest resonance mass and the continuum
threshold $t_c$, where one should note that some choices of $t_c$ in the sum rule literature do not satisfy this constraint. FESR is an useful complement of the previous sum rules. 
\vspace*{-0.25cm}
\subsubsection*{\b Gaussian sum  rule}
\nin
\beq
 { G}(s,\sigma)= \int_{t_<}^{\infty} dt~{\exp}^{-{(t+s)^2\over\sigma}}~{1\over\pi}{\rm Im} \Pi(t) \nnb
\vspace*{-.25cm}
\eeq
It is centered at $s$ with finite width resolution $\sqrt{4\pi\sigma}$. Mathematically, it can be used to 
derive the Borel/Laplace and FESR sum rules \cite{PEROTTET}. 
\vspace*{-0.25cm}
\subsubsection*{\b  $\tau$-like sum rule}
\nin
It has been originally introduced by \cite{BNP}:
\beq
R_{nm}(M^2_\tau)=\int_{0}^{M_\tau^2} dt~t^n\ga 1-{t\over {M_\tau^2}}\dr ^m ~{1\over\pi}{\rm Im} \Pi(t)~,
\vspace*{-.25cm}
\eeq
where $m=2$ for the physical $\tau\to \nu_\tau+hadrons$ decay. The threshold factor  suppresses the contribution near the real axis and improves the quality of the sum rule. This sum rule has been generalized in the literautre by the introduction of an arbitrary weight factor. 
\vspace*{-0.25cm}
\subsubsection*{\b  $\phi$-like and (pseudo)scalar sum rules}
\nin
In a similar way, a sum rule which suppresses the leading contribution threshold $t_c$ term has been inspired for extracting the stange quark mass from the $\phi$-meson sum rule \cite{SNI}:
\beq
R(t_c)=\int_{t_<}^{t_c} dt~\ga 1-2{t\over t_c}\dr ~{1\over\pi}{\rm Im} \Pi(t) ~,\vspace*{-.25cm}
\eeq
while a combination of (pseudo)scalar sum rules of the corresponding two-point correlator $\Psi_{(5)}(Q^2)$not sensitive to the leading perturbative (PT) contribution \cite{SNSU3}: 
\bea
&&\int_{t_<}^{\infty}{ dt\over t}~{\exp}^{-t\tau}~{1\over\pi}{\rm Im} \Psi_{(5)}(t)=\Psi_{(5)}(0)\nnb\\
&& +{3\over 4\pi^2}(\bar m_u\pm \bar m_s)^2\tau^{-1}\ga{\bar\alpha_s\over\pi}\dr+...,
\eea
has been introduced for extracting:
\beq
 \Psi_{(5)}(0)\equiv -(m_u\mp m_s)\la \bar uu\mp \bar ss\ra+{\rm Pert.~ terms},
 \eeq
 entering the Gell-Mann-Oakes-Renner relation.  
 \vspace*{-0.25cm}
\subsubsection*{\b  Double ratios of sum rules (DRSR)}
\nin
It is defined as:
\beq
r_{ij}={{\cal R}_i\over {\cal R}_j}~,
\eeq
for two channels $i$ and $j$ and is useful for extracting with a high accuracy the splittings of mesons
due to the vanishing of the $\alpha_s$ PT corrections and non-flavoured terms in the QCD expression of the sum rule~\cite{SNDRSR}.
\vspace*{-0.25cm}
\subsection{Parametrizations of the Spectral function}
\nin
\vspace*{-0.75cm}
\subsubsection*{\b Success and test of the na\"\i ve duality ansatz}
\nin
The spectral function Im $\Pi(t)$ can be measured inclusively from the data like $e^+e^-\to hadrons$, $\tau\to \nu_\tau+ hadrons$. In most cases, where hadron masses need to be predicted, there are no data available. Then, it is usual to use the minimal duality ansatz parametrization of the spectral function:
\beq
 One~ resonance \oplus{ QCD~ continuum}~\theta(t-t_c)~,
 \label{eq:ansatz}
\eeq
where $t_c$ is the continuum threshold and the QCD continuum comes from the discontinuity of the QCD diagrams in the OPE in order to ensure the matching of the two sides of the sum rules at high-energies \footnote{ In \cite{LUCHA} the continuum threshold $t_c$ is proposed to contain corrections of ${\cal O}(c_n/t_c^n)$ where $c_n$ are fitted from the data. However,  it is not clear that the two sides of the sum rules match at high energy. }.
The na\"\i ve duality anstaz in Eq. (\ref{eq:ansatz}) works quite well in the different applications of the SVZ sum rules 
within the corresponding expected 10-20\% accuracy of the approach. A test of this model from $e^+e^-\to hadrons$ and charmonium data have been done in \cite{SNB2} and has been quite satisfactory. Another successful test is the parametrization of the continuum of the pion spectral function by $3\pi$ final states using ChPT constraints  \cite{BIJP}, where the prediction for the  sum of light quark masses remains the same as the one  where a pion $\oplus$ a narrow $\pi(1300)$ is used despite the fact the $m_\pi^2$ is relatively light compared to the hadronic scale. 
\vspace*{-0.25cm}
\subsubsection*{\b Duality violation}
\nin
However, if one pushes the accuracy of the approach like, e.g., using $\tau$-decay data \cite{BNP,SNTAU},
one becomes sensitive to the detailed structure of the spectral function and a {\it duality violation} can affect the analysis. This problem is discussed in details in Shifman's talk~\cite{SHIFMAN,SHIF}.  
 In his talk, de Rafael \cite{RAFA} also discusses from a mathematical view, a large $N_c$ toy model (Von Mangoldt) for the spectral function which also leads 
to an oscillation behaviour and which is can be related to the previous model. 
Some phenomenological  implications of the model are discussed in \cite{SNTAU,PERIS,ALONSO}. 

 \vspace*{-0.5cm}
 \section{The original SVZ - Expansion}
\vspace*{-0.25cm}
 \nin
The SVZ ideas are not only related to the discovery of sum rules.  More important,  they have proposed a new way to phenomenologically parametrize (approximately) the non-perturbative (confinement) aspect of QCD beyond perturbation theory using an operator product expansion (OPE) \`a la Wilson in terms of the vacuum condensates of quark or/and gluons of higher and higher dimensions. \\
In addition to the well-known quark condensate $\la\bar\psi\psi\ra$ entering to the Gell-Mann-Oakes-Renner relation, SVZ have advocated the non-zero values of the gluon condensate $\la\alpha_s G^2\ra$ where they found to be about 0.04 GeV$^4$ from charmonium sum rules,  which has been verified in lattice simulations (see Zakharov's talk \cite{ZAK}) and from $e^+e^-$ data \cite{PEROTTET,LNT,YND}. 
More recent and refined analyzes indicate that its value is slightly higher of about $(0.06\pm 0.01)$ GeV$^4$ \cite{SNHeavy,SNI} which goes in line with the lower bound of about 0.08 GeV$^4$ obtained in \cite{BELL,BERTL} from heavy quark exponential moments.
The unconclusive extraction of $\la\alpha_s G^2\ra$ from a fit of $\tau$-decay data \cite{ALEPH} may be
due to the small value of the gluon condensate contribution in this channel where it acquires an extra-$\alpha_s$ correction, to its correlation with the other QCD parameters namely $\alpha_s$, $\la\bar\psi\psi\ra^2$ and an (eventual) quadratic $1/q^2$-term (see Zakharov's talk and next section).   
Indeed, the gluon condensates play an important r\^ole in gluodynamics (low-energy theorems, gluonia sum rules,...) and in some bag models as $\la\alpha_s G^2\ra$ is directly related to the vacuum energy density.
Taking the example of the Adler function in $e^+e^-\to hadrons$, it reads within the SVZ-expansion:
 \beq
   \vspace*{-0.15cm}
 -Q^2{d\over dQ^2}~\Pi(Q^2)
 =\sum_{d=0,1,2,...} {C_{d} \la 0| O_{d} |0\ra\over Q^{d}}
   \vspace*{-0.15cm}
\eeq
 The anatomy of the OPE up to $p=3$ is :
 \begin{itemize}
\small{ \item  {{ $d=0$~:}~~~~  usual PT series $ (a_s\equiv \alpha_s/\pi)$: \\
 $
 C_0= 1+a_s+1.640~ a_s^2+6.371~ a_s^3+49.076~a_s^4+$ \\
 \vspace*{0.1cm} 
 $
 ~~~{ \Delta_N\equiv \sum_{n>N} c_{n}\alpha_s^n~ }$ where $\Delta_N$ is unknown. 
}
\item  {   $d=2$~:}~~$\bar m_q^2$: small mass corrections 
\item  {  $d=4$~:}~~$\la \alpha_s ~G^2 \ra,~~m_q\la\bar \psi_q\psi_q\ra$: \\ gluon and quark condensates
\item  {   $d=6$~:}~~ $\alpha_s\la\bar \psi_q\psi_q\ra^2$, $gf_{abc}\la G^aG^bG^c\ra,$ $ m_q g\la \bar\psi\lambda^a\sigma_{\mu\nu}G^{\mu\nu}_a\ra$~~: four-quark, triple gluon and mixed  condensates,
}
\end{itemize}
where one should notiice that in this channel the coefficient of the triple gluon condensate vanishes in the chiral limit $m_q=0$.
\vspace*{-0.5cm}
\section{Theoretical progresses}
\vspace*{-0.25cm}
\nin
There have been different steps for improving the original SVZ sum rules.
\vspace*{-0.4cm}
\subsection{Optimization criteria from quantum mechanics}
\nin
\vspace*{-0.25cm}
\begin{center}
\vspace*{-1cm}
{\begin{figure}[hbt]
{\includegraphics[width=7cm]{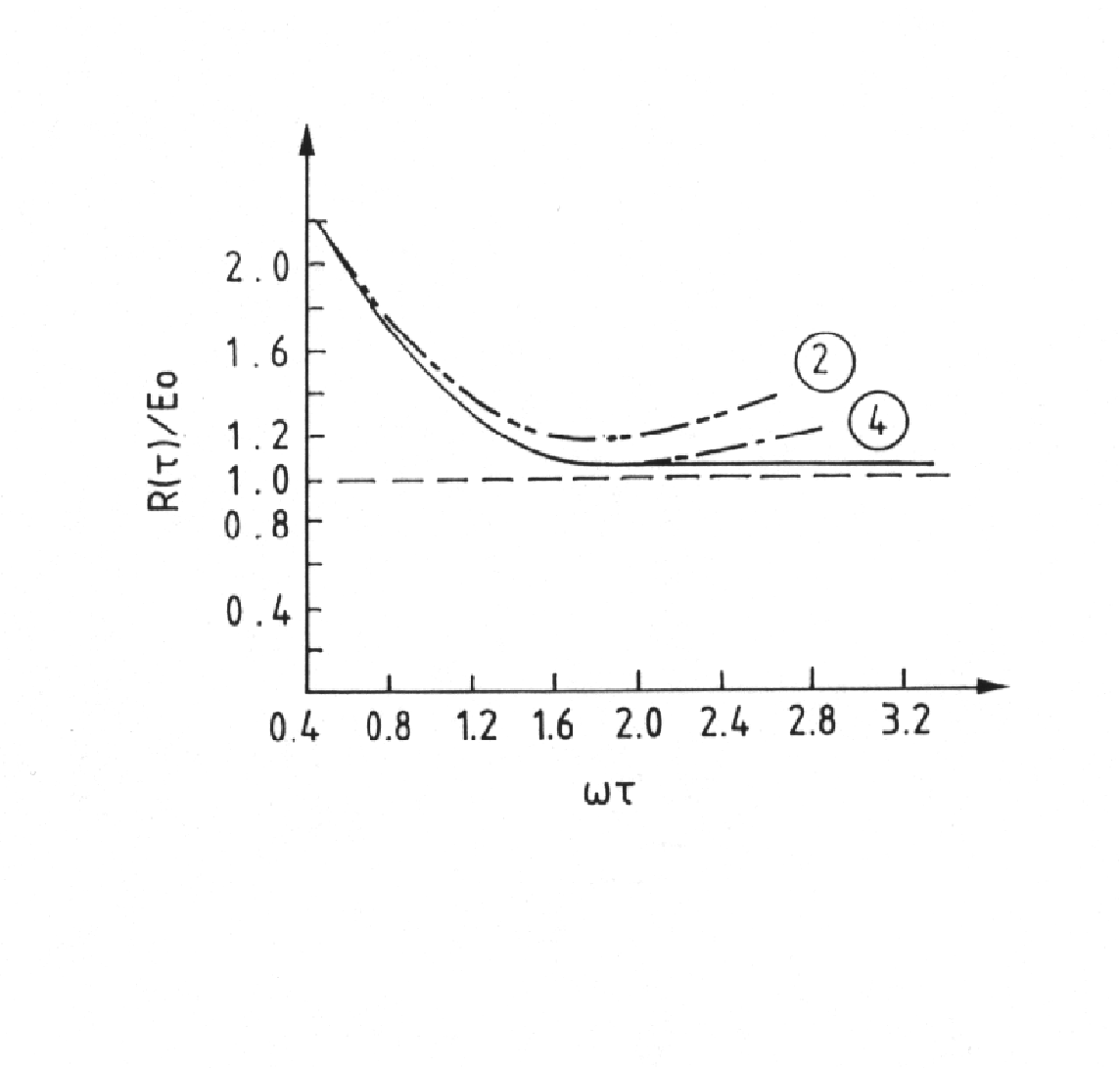}}
\vspace*{-0.25cm}
{\includegraphics[width=6cm]{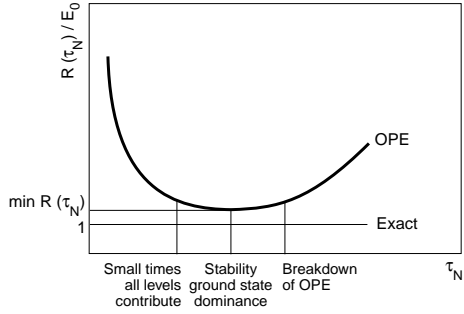}}
\caption{Ratio of moments ${\cal R}(\tau)$ in Eq. (\ref{eq:ratio}) versus the sum rule variable $\tau$: a) harmonic oscillator. b) charmonium.}
\label{fig:oscillator}
\end{figure}}
\end{center}
The question to ask is how one can extract an optimal information on resonance properties from the sum rules and in the same time the OPE remains convergent. The original SVZ proposal is to find a window where the resonance contribution is bigger than the QCD continuum one and where the non-perurbative terms remain reasonnable corrections. Numerically, the argument is handwaving as the percent of contribution to be fixed is arbitrary. 
In a series of papers, Bell and Bertlmann \cite{BERTL,BELL} have investigated this problem using the harmonic oscillator within the exponential moment sum rules. The sum rule variable $\tau$  is here an imaginary time variable. The analysis of the ratio of moments ${\cal R}(\tau)$ defined in Eq. (\ref{eq:ratio}) is
shown in Fig. \ref{fig:oscillator}a), where one can observe that the exact solution (ground state energy $E_0$) is reached when more and more terms of the series are added and the optimal information is reached at the minimum 
of $\tau$ for a truncated series. The position of this minimum co\"\i c\i nde with the SVZ sum rule window but more rigourous. For a comparison, the case of the $J/\psi$ mass is shown in Fig. \ref{fig:oscillator}b). \\
Another free parameter in the phenomenological analysis is the value of the continuum threshold $t_c$. Many authors {\it adjust its value} at the intuitive mass of the next radial hadron excitation. This procedure can be false as the QCD continuum only smears all high mass radial excitations, and what is important is the area in the sum rule integral. As the value of $t_c$ is like the sum rule variable an external parameter, one can also require that the 
physical observables (the lowest resonance parameters) are insensitive to its change. In some cases, this $t_c$ stability is not reached due to the simplest form of the ansatz in Eq. (\ref{eq:ansatz}). In this case, the complementary use of FESR is useful due to the correlation of $t_c$ with the mass of the lowest resonance  (but not with the one of the radial excitation)  \cite{PEROTTET}. 
\vspace*{-0.25cm}
\subsection{Renormalizations and radiative corrections}
\nin
SVZ original works have been done to lowest order in $\alpha_s$. There have been intensive activities for improving the SVZ during the period of 80-90 :
\begin{itemize}
\small{
\item Inclusion of the PT $\alpha_s$ corrections to the exponential sum rules reveals  inverse Laplace transform properties rather than a Borel one \cite{PSEUDO}.
\item Mixing of operators under renormalization and evaluation of their anomalous dimensions give a more precise meaning of the condensates where some combinations have been found to be renormalization group invariant \cite{TARRACH,BNP}.
\item Absorption of the light quark mass singularities into the condensates  leads to the definition of normal  or non-normal ordered quark condensate \cite{PSEUDO,BROAD,BNP}. 
\item Evaluation of the contributions of high-dimension condensates for testing the convergence of the OPE \cite{95}.
\item Evaluation of the higher order PT corrections \cite{CHETYRKIN,KORNER} and of the ones of the Wilson coefficients of the condensates in some channels \cite{PASCUAL}. 
}
\end{itemize}
\vspace*{-0.25cm}
However, despite these large amounts of efforts in the past, it is disappointing to note that most of the recent applications of QSSR only limit to the LO in $\alpha_s$. 
\vspace*{-0.5cm}
\section{Traditional QSSR phenomenology}
\vspace*{-0.25cm}
\nin
Since the original work of SVZ \cite{SVZ}, the rich conventional phenomenology of QSSR has been reviewed in  \cite{DOSCH,CONF,ZAKA,BERTL,RRY,NOV,SHUR,IOFFE,REV,SNB1,SNB2}. The different talks given in this session indicate the continuous and wide range of activities in this field. I flash below a panorama of its impressive applications in hadron physics \footnote{The approach has been also applied to other QCD-like models like composite models \cite{WZB} and supersymmetry \cite{SQCD}.}:
\begin{itemize}
\small{
\item $\rho$ meson, gluon condensate, charm mass { \footnotesize since 1979 \cite{SVZ}}
 \item Meson spectroscopy {  \footnotesize since 1981 \cite{RRY}}
\item Light quark masses {  \footnotesize since  1981 \cite{PSEUDO,SCAL}}
\item Corrections to $\pi$ and $K$ PCAC  since 1981 {  \footnotesize \cite{SNSU3,SNB2,PSEU}}
\item Heavy quark masses  \footnotesize { since 1979 \cite{SVZ, SN1, SNBC3,MARQ, ZYAB,SN10} } 
\item Condensates \footnotesize { since 1979 \cite{SVZ, RRY, ZAL,YND,SNHeavy,ZYAB,SN10,MENES,LNT,PEROTTET,SNI,ALEPH} } 
\item Heavy-light mesons {  \footnotesize since 1978 \cite{102,SNFBI}}
\small \item Light baryons  {  \footnotesize since 1981 \cite{DOSCH,IOFFEBAR,DOSCHBAR}}
  \item Heavy baryons  {  \footnotesize since 1992 \cite{CHABAB}}
 \item Gluonium {  \footnotesize since 1981 \cite{NSVZ,NGLUE}}
   \item Light hybrids {  \footnotesize since 1987 \cite{211}}
 \item Heavy hybrids {  \footnotesize   since 1985 \cite{217,SNB2}}
     \item Four-quarks, molecules {  \footnotesize since 1985 \cite{219,NIELSEN} }
  \item Hadronic decays since 84: { \footnotesize Vertex \cite{PAVERN,NAVA},  Light Cone \cite{FAZIO}}
 \item $\tau$-decay since 1988 { \footnotesize \cite{BNP,LEDI,SNTAU,VALENZ}}
 \item Thermal and in-medium hadrons since 1986 { \footnotesize \cite{BOCH,LOEWE}}~.
  }
\end{itemize}
In most applications, the sum rule predictions using the optimization criteria are successful compared with data when available or with some other non-perturbative approaches.
\vspace*{-0.5cm}
\section{Large order and quadratic terms in PT theory}
\vspace*{-0.25cm}
\nin
Besides the question of duality violation for the spectral function, it is equally important to control the large order terms of the QCD perturbative series. This issue is discussed in details in the talk of Zakharov \cite{ZAK} which we only sketch here. 
One attempts to answer the nature of the reminder of the PT series for large $N$:
\beq
\Delta_N\equiv \sum_{n>N} c_{n}\alpha_s^n
\eeq
 \vspace*{-0.25cm}
 \subsection{Renormalons and gluon condensate}
 \nin
 According to the usual wisdom, the coefficient $c_n$ of series is expected to grow like $n!$ asymptotically (Infrared Renormalon), which is expected to be absorbed into the perturbative part of the gluon condensate $\la\alpha_s G^2\ra$, while the uncertainty of the asymptotic series would induce a term proportionnal to $\Lambda^4$ which is the {\it physical} gluon condensate  appearing in the OPE.  
Moreover, the full value of the gluon condensate can be measured with high-accuracy from the lattice
as it is the plaquette action. \footnote{Some pioneer attempt to measure the gluon condnesate on the lattice can be found in \cite{GIACO}.} It can be expressed as :
\beq
\Delta P_N\equiv P_{exact}-\sum_N{p_n\alpha_s^n}\approx (\Lambda\cdot a)^{\rho_N}
\eeq
where $P_{exact}$ is the exact plaquette action, $p_n$ is the perturbative coefficient calculated explicitly and $a$ is the lattice spacing. Fitting the lattice data on the difference $\Delta P_N$ \cite{RAKOV} using power-like corrections, one finds:
\bea
\rho_N \simeq&& 2 ~~~for ~~~N\leq 10\nnb\\
&&4 ~~~for ~~~N\geq 10~.
\eea
The previous result indicates  that numerically the PT series can differentiate between the quadratic and quartic corrections and shows some duality between long perturbative series and power corrections. Another remarkable feature from lattice calculations is the fact that:
\beq
r_n\equiv {p_{n+1}\over p_n}\approx constant~,
\eeq
indicating that the series grow geometrically. Some other QCD processes present analogous properties \cite{SNZ}. These results indicate no alternate signs for the terms of the QCD PT series, which then do not support (to the order where the series are evaluated) the presence of the Ultraviolet Renormalon. 
They  also  indicate that the PT series do not show any existence of Infrared Renormalon (no $n!$ growth of the PT series) at that order.  All these features can change our dogma on the understanding of higher order perturbative series. 
\vspace*{-0.25cm}
\subsection{Quadratic corrections}
\nin
A similar observation of duality between the long PT series and quadratic corrections has been reported in the analysis of the Coulomb potential where the quadratic correction at short distance:
 \beq
 V_{\bar QQ}\vert_{non-pert} =\sigma\cdot R~,
 \eeq
($R$ is the distance between two heavy quarks and $\sigma$ is the string tension) is reproduced by higher order PT series \cite{SOMMER}. Another similar decrease of the strength of the quadratic correction when adding higher order PT terms has been also noted in the  neutrino DIS analysis of the $xF_3$ sum rule \cite{KATAEV}, where in this process, the quadratic corrections are instead associated to long distances \footnote{
Quadratic corrections also to play a r\^ole in the Regge trajectories of mesons \cite{MEGIAS}.}. \\
{\it What seems to emerge from the previous observations is the Duality between Long perturbative series and Quadratic corrections: one can use one of them but not both in order to avoid a double counting.} \\
However, as there is no gauge invariant operator of dimension 2 in a field theoretical approach, it is difficult to parametrize analytically a such quadratic term.  In fact, from the previous discussion, the quadratic term is a part of the Wilson coefficient of the unit operator which is difficult (in practice) to disentangle from the other terms of the  PT series poorly known to $N\leq 3\sim 4$. Some possible issue for explaining the origin of the quadratic term is the dual string model or the AdS/QCD approach~\cite{ZAK}.
\vspace*{-0.5cm}
\section{Phenomenology of Quadratic corrections}
\nin
\vspace*{-1cm}
\subsection{Short-distance gluon mass}
\nin
As emphasized in previous section, it is difficult to parametrize analytically the contribution of 
the quadratic correction despite its evidence from numerical simulations. An attempt to include
this contribution in a gauge invariant way (to leading order in $\alpha_s$) is to introduce a tachyonic 
gluon mass through the gluon propagator \cite{CNZ}:
\beq
{1\over q^2}~~\to~~ {1\over q^2+\lambda^2}
\eeq
A systematic evaluation of this contribution has been presented in \cite{CNZ} for different two-point correlators. 
\vspace*{-0.25cm}
 \subsection{Correlated estimate of $\lambda^2$}
 \nin
 The value of $\lambda^2$
has been fitted \footnote{More detailed discussions can be found in  \cite{SN05,SNTAU}.}
from 
$e^+e^-\to hadrons$ data \cite{SNI,CNZ}, $\pi$-Laplace sum rule \cite{CNZ}, and lattice data for the sum of pseudoscalar $\oplus$ scalar two-point correlators \cite{SNZ0}, where some eventual instanton contributions cancel out. We show the results in Table~\ref{tab:tachyon}.
One can note that the one from $\tau$-decay  \cite{SNTAU} has been estimated from the difference between the large $\beta$-limit prediction and the sum of the known PT series which then depends on how the PT series is resummed (fixed order or contour improved). 

{\footnotesize            
\begin{center}
\footnotesize            
{\begin{table}
\setlength{\tabcolsep}{1.25pc}
\caption{Estimate of  $d_2\equiv { -(\alpha_s/\pi)} \lambda^2 $ in GeV$^2\times 10^2$ from light quarks }
\label{tab:tachyon}
\begin{tabular}{lll}
\hline
 \footnotesize             Channels&  \footnotesize            &  \footnotesize       $d_2$      
\\
\hline
\footnotesize            \boldmath$e^+e^-$ {\bf data}& \\
\footnotesize  {\it Ratio of moments ${\cal R}(\tau)$}&& \footnotesize$6.5\pm 0.5$\\
\footnotesize            \boldmath $ \pi-$ \bf sum rule& \\
\footnotesize{\it Laplace ${\cal L}(\tau)$}&&  \footnotesize $12\pm 6$\\
\footnotesize{\bf Lattice data}&& \\
\footnotesize{\it Scalar+Pseudoscalar}&& \footnotesize $\simeq 12$\\

\footnotesize          \bf \boldmath   $\tau$-decay &\\
\footnotesize  $\Delta_N\equiv~large~ \beta-\sum_1^4$ PT series :& \\
\footnotesize             Fixed Order PT &&\footnotesize $2.6\pm 0.8$\\
\footnotesize             Contour Improved &&\footnotesize $5.9\pm 0.8$\\
\footnotesize \it \bf Arithmetic Average& & \footnotesize\boldmath$  7\pm 3$ \\
\hline
\end{tabular}
\end{table}
}
\end{center}
\vspace*{-1.cm}
 }
 \vspace*{-0.25cm}
 \subsection{Effects of $\lambda^2$ on the  light quark masses}
 \nin
 The sum rule scale of the $\pi$-channel has been puzzling in the conventional
 approach where we have the hierarchy \cite{NSVZ}:
 \beq
 M^2_\rho\approx 0.6~{\rm GeV^2}\ll M^2_\pi\approx 2.7~{\rm GeV^2}~, 
 \eeq
while in the presence of $\lambda^2$:
 \beq
 M^2_\rho\approx M^2_\pi\approx 1.7~{\rm GeV^2}~, 
 \eeq
where the duality between the two sides of the $\pi$-sum rule improves. $\lambda^2$ decreases the
 estimate of the light quark masses by 5\% from the (pseudo)scalar sum rules \cite{CNZ}. 
 The extractions of the strange quark masses from $e^+e^-$ and $\tau$-decay data including
 the $\lambda^2$ corrections have been done in \cite{SNms,SNTAU}. The averages of the results
from different forms of the sum rules are in MeV:
\bea
\overline m_u(2)&=&2.8(2)~,~~~\overline m_d(2)=5.1(2)~,\nnb\\
\overline m_s(2)&=&96.1(4.8)~.
\eea
where the running masses have been evaluated at 2 GeV. 
 \begin{center}
\footnotesize          
{\begin{table}[hbt]
\setlength{\tabcolsep}{0.8pc}
\caption{Different QCD corrections to the $\tau$ hadronic widths.}
\label{tab:correct}  
\begin{tabular}{lll}
\hline
     \footnotesize             Corrections&      \footnotesize            &      \footnotesize             Size $\times 10^3$ 
\\
\hline
$\delta_{svz}=\sum_4^8\delta^{(D)}$&&  \footnotesize $-(7.8\pm 1.0)$\\
    \it $\delta_{st}\equiv \delta_{svz}+\delta_m^{(2)}+\delta_{\pi}+\delta_{a_0}$&&      \footnotesize \it $-(10.9\pm 1.1)$ \\
$\delta_{inst}$&& \footnotesize            $-(0.7\pm 2.7)/20$\\
$\delta_{DV}$&&  \footnotesize $-(15\pm 9)$ \\
  \footnotesize$\delta_{2}\equiv  large~ \beta-\sum_1^4 PT$&&   \footnotesize $(17\pm 5)$ FO\\
&&  \footnotesize $(39\pm 5)$ CI \\
    \it $\delta_{nst}\equiv \delta_{inst}+\delta_{DV}+\delta_{2}$&&     \footnotesize  \it $(2.0\pm 9.4)$~~~ FO\\
&&    \footnotesize   \it $(24.0\pm 10.6)$~~~ CI \\
\hline
\end{tabular}
\end{table}
}
\end{center}
\nin
\vspace*{-1.5cm}
 \subsection{Effects of $\lambda^2$ on $ \alpha_s$ from $ \tau$-decays}
 \nin
 The $\tau\to \nu_\tau+hadrons$ process is a good laboratory for testing the effects of
these tiny  deviations (duality violation and tachyonic gluon mass) from the conventional sum rules approach. The non-strange $\Delta S=0$ component of the $\tau$-hadronic width normalized to the electronic width can be expressed as:
\bea
R_{\tau} 
  &=&  3 |V_{ud}|^2 S_{EW}\times\nnb\\
 && (1+\delta^{(0)} + \delta'_{EW}+ \delta_m^{(2)}+ \delta_{\rm svz}+\delta_{\rm nst})~,
\eea
where   $|V_{ud}| =  0.97418  \pm \, 0.00027$ \cite{PDG} is the CKM mixing angle;
$
S_{EW} = 1.0198 \pm 0.0006$
\cite{MARCIANO} 
{\rm and} $ \delta'_{EW} = 0.001 $
\cite{LI}. 
are known electroweak corrections; $\delta^{(0)}$ and $\delta_m^{(2)}$ are the perturbative and light quark mass corrections; 
$
\delta_{\rm svz}\equiv \sum_{D=4}^8\delta^{(D)}~,
$
is the sum of the non-perturbative (NP) contributions of dimension $D$ within the SVZ expansion \cite{SVZ}, while $\delta_{\rm nst}$ are some eventual NP effects not included into $\delta_{\rm svz}$, which are here due to instanton ($\delta_{inst}$), quadratic corrections ($\delta_{2}$) and duality violations ($\delta_{DV}$). The size of different contributions are given in Table \ref{tab:correct}, where it is remarkable to note that the contributions of duality violation and quadratic corrections tend to cancel out.   Using: 
$
R_{\tau,V+A}\vert_{exp}=3.479\pm 0.011
$, we deduce the result in Table \ref{tab:alphas}, which we compare with some other determinations. For completing, the previous values of the QCD perturbative
parameters, we give the recent values of the $c,~b$ running quark masses from ratio of moments to order $\alpha_s^3$ \cite{SN10}
in units of MeV:
\beq
\overline m_c(m_c)=1261(18)~,~~~~~\overline m_b(m_b)=4232(10)~.
\eeq
\vspace*{-0.5cm}
\begin{table}[hbt]
\setlength{\tabcolsep}{0.8pc}
\footnotesize   
\caption{Value of $\alpha_s$ rom $\tau$-decay and comparison with the one from $Z\to hadrons$ \cite{DAVIER}   
and the world average \cite{PDG,BETHKE}}     
\label{tab:alphas}
\begin{tabular}{llll}
\hline
\footnotesize PT&$\alpha_s(M_\tau)$&$\alpha_s(M_Z)$&\\
\hline
\footnotesize FO &$0.3276~(34)_{\rm ex}(86)_{\rm th}$&$0.1195~(4)_{\rm ex}(10)_{\rm th}(2)_{\rm ev}$&\\
\footnotesize CI & $0.3221~(48)_{\rm ex}(122)_{\rm th}$&$0.1188~(6)_{\rm ex}(15)_{\rm th}(2)_{\rm ev}$&\\
\footnotesize$\la~\ra$&$0.3249~(29)_{\rm ex}(75)_{\rm th}$&$0.1192~(4)_{\rm ex}(9)_{\rm th}(2)_{\rm ev}$&
\\
\hline
\footnotesize  Z&&  $0.1191~(27)_{\rm ex}(2)_{\rm th}$&\\
\footnotesize $\la~\ra$&&  0.1184~(7)&\\
\hline
\end{tabular}
\end{table}
We  also show in Table \ref{tab:param} the values of non-pertubative condensates determined from QSSR. 
\vspace*{-0.5cm}
{\scriptsize
\begin{center}
\begin{table}[hbt]
\setlength{\tabcolsep}{0.5pc}
 \caption{\scriptsize  Values of the QCD condensates from QSSR. }
 \label{tab:param}
\begin{tabular}{lll}
\hline 
\footnotesize  Condensates &\footnotesize Values [GeV]$^{   d}$&\footnotesize Sources\\
\hline
   $\la\bar uu\ra(2)$&\footnotesize$-(0.254\pm .015)^3$&\footnotesize   (pseudo)scal,
\\

$\la \bar dd\ra/\la\bar uu\ra$&\footnotesize$1-9\times 10^{-3}$&\footnotesize non-norm. ord.   (pseudo)scal, \\
$\la \bar ss\ra/\la\bar dd\ra$&\footnotesize$0.74(3)$&\footnotesize non-norm. ord.  (pseudo)scal\\
&&\footnotesize   $\oplus$ light \& heavy baryons\\
 $\la\alpha_s G^2\ra$&\footnotesize$7(2)10^{-2}$ &\footnotesize   $e^+e^-,~\Upsilon-\eta_b,~J/\psi$ Laplace\\
&&\footnotesize   $\tau$, $J/\psi$ mom : unconclusive\\
  $g\la \bar\psi G\psi\ra $&\footnotesize$M^2_0=0.80(2)$ &\footnotesize   Light baryons, $B,~B^*$\\
 $g^3\la G^3\ra$&\footnotesize$(31\pm 13)\la \alpha_s G^2\ra$&\footnotesize    $J/\psi$-mom 
 \\
$\rho\alpha_s\la\bar\psi\psi\ra^2$&\footnotesize$(4.5\pm 0.3)10^{-4}$ &\footnotesize   $\lrar               \rho=2.1\pm 0.2$\\
\hline
\end{tabular}
\end{table}
\end{center}
 }
 \vspace*{-1.35cm}
\section*{Conclusions}
\vspace*{-0.35cm}
\nin
We have presented in a compact form the developments of the SVZ sum rules 30 $\oplus$ 1 years later
and have summarized in different Tables the values of the QCD parameters from the approach. Its applications are rich and have the advantage (compared to lattice simulations) to be analytical. The successful applications of the sum rules motivate continuous phenomenological applications, and improvements of the approach in connection with its new developments discussed here and by SZ in this jubilee, namely the inclusion of radiative and quadratic corrections, the study of duality violation and the connection of QSSR to dual AdS/QCD models. 
 \vspace*{-0.5cm}
\section*{Acknowledgements} 
\vspace*{-0.25cm}
\nin
I wish to thank my numerous collaborators and especially Valya Zakharov for continuously stimulating my  interest in this field.
\vspace*{-1cm}

\end{document}